# A Cloud collaborative approach for managing patients wellness

Angel Ruiz-Zafra, Manuel Noguera, Kawtar Benghazi,
José Luis Garrido
Departamento de Lenguajes y Sist. Informáticos, Granada, Spain
{angelr,mnoguera,benghazi, jgarrido}@ugr.es

Sergio F. Ochoa
Computer Science Department, Universidad de Chile, Chile
sochoa@dcc.uchile.cl

*Abstract*—Patients with chronic diseases or people with special health care needs are typically monitored by various health experts that address the problem from several perspectives. These experts usually do not interact directly between them; therefore, the instructions given to the patient by one of them are provided disregarding advices and instructions provided by the others. The collaboration between different health experts in a real context is mandatory to ensure the proper monitoring of the patient. This kind of collaboration, supported by technology, benefits the users' health condition and helps the patient achieve their goals in terms of wellbeing. This paper presents an approach for collaborative management of events related to activity supervisions and monitoring of these patients types. It also introduces a model and a mobile application to support this collaborative work.

## I. INTRODUCTION

In the last few years there has been a change of mind about the importance of health and wellness. This change of mind has raised a demand of tools and solutions to help people to lead a healthier life in different aspects such as nutrition, physiotherapy and physical activity [1-3].

The emergence of mobile devices, such as smartphones, and powerful computer paradigms, such as cloud computing, has resulted in useful systems in many application domains. The attention paid to wellness and healthcare, together with technological innovation, has given birth to a new market of solutions, and several research avenues in each domain related to wellness and healthcare (e.g., nutrition, physical activity) [4-5].

Integral solutions (that involve software and hardware) range from wearable devices to measure physical and physiological variables [6], to smartphone applications consisting of diary or tracker functionalities [7]. Usually, these software solutions are focused on recording a specific aspect of the people, such as their physical activity, nutrition or physiological variables (e.g., heart rate, temperature, blood pressure). Typically, they are designed to be used in standalone user settings. Some other more sophisticated systems allow managing, monitoring and even approximating the physical condition of the user on the basis of such records [7].

In the medical context, the interdisciplinary collaboration between experts from different domains (e.g., nutrition, psychologists and physicians) is a fundamental requirement for the supervision of various patients types; for instance athletes, or people suffering chronic diseases.

To achieve this collaboration, the experts should have daily, weekly or monthly meetings to exchange information, points of view and feedback from the patients. These meetings require that they have to adjust their schedules and move to the meeting point, which is expensive (mainly in terms of time) and jeopardizes the effective attendance of the people to these meetings.

However, this interdisciplinary collaboration of experts is anyway a crucial part in the patient monitoring process. The possibility to provide a consensual feedback enables the patient to have a full and customized monitoring solution. In this way, the patient healthcare condition is supervised by a group of experts in different domains who are aware of the decisions of the rest of experts. This interdisciplinary monitoring can take advantage of the advances in ICT, which open endless possibilities to design and develop novel solutions where the feedback from the experts to the patients (and vice versa) can be at real-time, and positively impact health condition of these people.

The use of a novel technology by itself does not ensure a proper monitoring or improve the health condition of the people. Conversely, it is a prior concern the design and use of a coordination and collaboration model that enables suitable interactions between experts (at real-time, remotely, anywhere and at anytime), as well as the architecture to support it.

In this paper, we propose an approach for collaboratively managing the supervision and treatment process of these patient types. This approach should be considered when designing healthcare systems where experts from different domains and end-users (e.g., athletes, patients with chronic diseases, or people on diet) are involved. The solution presented in this paper proposes a collaborative work model to represent collaboration scenarios in this domain, as well as the architecture to support them.

The rest of the paper is organized as follows. Section II discusses the related work. Section III presents the proposed collaboration model. Section IV briefly describes a mobile application that supports the patients monitoring and expert's coordination activities. Section V presents the conclusions and future work.

## II. RELATED WORK

Several collaborative software solutions have appeared in different domains based on different technologies. Originally, the first systems for supporting cooperative work were designed for addressing office work [8] and they were supported by basic technology, such as emails, chats and discussion boards. Over time, the collaborative systems have evolved in parallel with the technology and the needs of people working in several domains. Because of this, various solutions have appeared in domains such as videogames [9] and e-learning [10].

Many software solutions have also appeared to support collaboration in several healthcare domains, in order to provide patients, athletes or other kind of end-users, a tool to perform their activities or helping them in the health treatments. For example, the software presented in [7] can be used as a tracker by athletes; and the system presented in [11] allows people to get their heart rate information and validate it.

The main disadvantage of these systems is that users have to use the applications on their own, without any supervision of an expert in the health care domain. In order to address this disadvantage, the use of collaborative solutions is highly recommended.

It is worth mentioning that there are some self-managing systems, which involve patients and remote experts as users. For instance, a collaborative game that intends to change the behavior in nutrition of the people [12], a tele-rehabilitation system for brain-injured patients [13], or a mobile healthcare system for wheelchairs users and medical assistance support [14]. These collaborative systems involve patients and experts, and they are designed to work in specific healthcare domains to cover a specific problem.

In this paper we propose a collaborative model to cover the collaboration requirements between different experts from different fields, as well as a technological solution (architecture) to support this collaborative model. Next section describes this model.

## III. COLLABORATIVE SUPERVISION MODEL

### A. Collaboration in the wellbeing domain

In wellbeing domain, different disciplines are involved, and the collaboration among specialists is required to get an integral treatment of the patients. This collaboration is traditionally conducted through daily, monthly or weekly meetings, where the experts exchange points of view, opinions and feedback about the patient. This traditional approach is limited, due it imposes strong restrictions to the involved people to coordinate activities, exchange information or make decisions about how to adjust a treatment.

The use of novel computer paradigms, as well as mobile technology and wireless communication, enables the collaboration through the cloud, at anytime, anywhere and remotely via smartphones. The systems that intends to take advantage of these new technologies should also consider two additional concerns to address the user satisfaction and the improvement of the patients wellness condition:

- *Collaborative model*: How the collaboration between experts, and between experts and users, should be managed to provide good results, and which features should be considered to support as many scenarios as possible.
- *System architecture*: This considers the hardware and software solutions used to support the collaborative scenario. A robust, scalable and extensible architecture is mandatory to increase the range of healthcare scenarios that can be addressed.

These concerns underlay in the design of collaboration scenarios for this domain.

### B. Collaboration Scenarios

In the wellness domain there are concerns to be considered in order to ensure quality of service:

- *User care*: The continuous supervision of patients over time builds trust and wellbeing in these end-users, which directly affects their health condition. The continuous monitoring provided by experts' feedback, enables enhanced user care.
- *Meet the healthcare outcomes*: Leading to results, which show the evolution of the user in terms of health condition.

After several meetings with researchers from the wellbeing domain, various collaboration scenarios were identified, which are depicted in Figure 1.

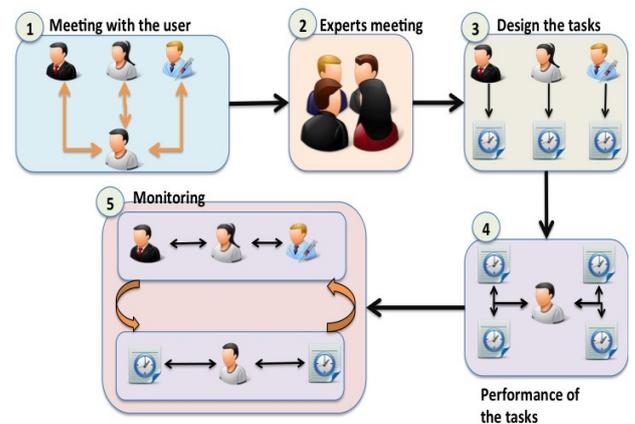

Fig. 1. Collaboration scenarios in the wellbeing monitoring

The figure represents the general interactions contexts among the involved people, where experts from different areas and end-users (i.e., patients or monitored people) participate. These collaborative scenarios comprise the following situations:

1) Each expert in an domain has a face-to-face meeting with the end-user (i.e., a patient) to evaluate him/her and get feedback.

2) Each expert shares (according to the previous feedback) relevant health information with the rest of experts. For instance, in case of an athlete, if a physician detects that the patient has cardiovascular diseasehis will be considered by the coach to prepare the training plan or the nutritionist to prepare the diet.

TABLE I. INTERACTION MAP CONSIDERED IN THE MODEL

| Notif. Type | Interaction | Description |
|---|---|---|
| Type 1 | Expert ←→ Expert | One expert reports to the rest of experts a decision taken on his/her field about the activities/evolution of the patient. |
| Type 2 | Expert ←→ Expert | One expert reports to the rest of experts a decision taken that should be accepted by the rest of experts, before being sent to the patient because other domains are involved or other experts should be informed. |
| Type 3 | Expert → User | One expert changes the patient activity (e.g., nutrition, exercises, etc.) and the latter is notified. |
| Type 4 | Expert → User | One expert changes the patient activity and the latter is not notified. |
| Type 5 | User → Expert | The user notifies the expert(s) about the evolution of the activities performed and the results. |
| Type 6 | User → Expert | The patient notifies to the expert(s) about a goal reached. |

3) Each expert, according to the information of the rest of experts and his/her own expertise, defines the goals and timetables for the patient.
4) The patient starts with the different tasks assigned by the experts; for instance: physical activity, diet, physiological exercise, or scheduled medical reviews.
5) Experts interact between them and with the patients, and the patients with the experts, in order to change the opinions about goals, schedules and outcomes of a treatment.

Besides these concerns, the following features must be met by systems that support these collaborative scenarios:

- *Real-time communication support*: Unlike other systems, where the response time is not important (such as emails or document exchange), the possibility to provide a consensual advice in real-time is mandatory to ensure the people monitoring (user care) correctly.
- *Two types of users (at least)*: Supervisors/experts from the different domains (e.g., nutrition, psychology, physician, coach, etc.) and patients.
- *Three types of interactions*: It is important that the system supports full communication and interaction between these users types: experts→experts, expert→user, and user→experts.
- *Different types of notifications*: This is required because there are different types of interactions; therefore, it is important to design a solution that enables the possibility to support different types of notifications to cover as many situations as possible. A notification is a real-time interaction between experts and/or patients.
- *Tasks*: A user who wants to improve his/her healthcare condition in some aspect usually should follow a set of instructions defined by his/her supervisor(s); this set of instructions is a task. The tasks could consist of performing some exercise to lose some weight or to perform rehabilitation, and also monitoring their heart rate of a patient.

C. *Collaborative-work Model*

Collaborative work encompasses several activities. For instance to exchange documentation, the collection and process of the information produced in the meetings, or the coordination of meetings per day/week/month. Some classical approaches have addressed the coordination between experts on the basis of shared workspaces through which to plan meetings and exchange documents.

Mobile technologies allow meetings to be planed and documents to be exchanged in a more agile manner from personal devices. Furthermore, one of the main requirements for using computer-supported collaborative work in healthcare domain, is the interaction between experts/supervisors when they have to make important decisions. In this area, a change in the patient behavior or habits in a specific field can affect other aspects of his/her treatment, in terms of the tasks or goals. For example, if the nutritionist changes the diet of an athlete, a coach should know it, in order to adjust the training according to the diet, and a physician should approve the new training plan to ensure that this is not a problem for the athlete health condition.

In order to achieve this, a solution based on the use of notification (a real-time message, alert, etc.) is proposed. The Figure 2 shows a model to represent the notifications and the types of users, as well as the possibility to add new types of notifications.

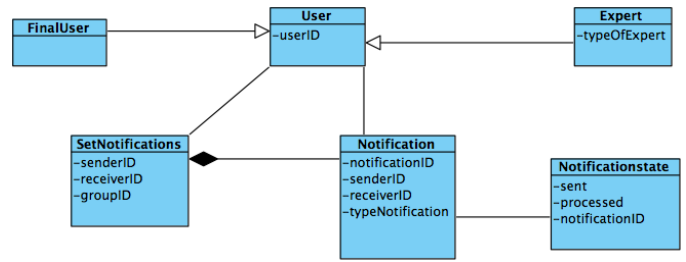

Fig. 2. Notification model for supporting collaboration

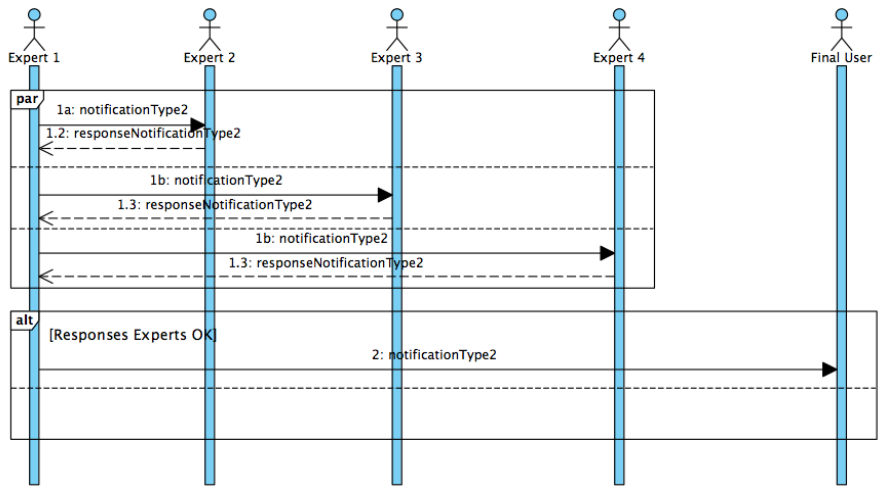

Fig. 3. Example of notification sequence

Different types of notifications to address the various concerns that are present in the wellness domain have been identified. Table 1 summarizes, in terms of interactions types, the notifications considered by the model.

Figure 3 shows an example of a sequence of interactions between experts and a user, where one expert sends a notification to the user that requires the approval of the rest of the experts.

An expert who wants to send a notification (Type 2) to the final user (i.e., the patient), but the remaining experts confirmation is required. At the same time, Expert 1 can send the rest of experts the notification (shown through a pop-up message for example) and wait for the responses. If all of them respond "OK", the system automatically sends the notification to the end-user.

This entails that changes that are occurring in the planning of the patient treatment, are transparent to the latter, because the experts collaborate between them through the cloud.

### D. Architecture for Notification Management

Mobile technology provides useful capabilities to support the collaborative work, such as easy communication between devices, delocalized interaction, transportability and computing capability that allows perform activities/tasks in an easy way. This can be combined with cloud computing services that provides scalability, robustness, security and centralized information, among others. The architecture proposed for supporting the model is based on these two technologies.

The use of these technologies allows developing solutions to implement the proposed collaborative work model, providing also features such as:

- *Real-time*: Push/pull technology is, nowadays, widely used in smartphones and well supported by cloud technology. Using push/pull technology it is possible to send and receive customized notification in real-time.
- *Scalability*: It is possible to implement the collaborative scenarios proposed with an unknown number of experts and end-users. Therefore, the collaborative work model, supported by the proposed cloud-based architecture and applied in a custom solution, ensures that the resulting system will have a suitable response time, regardless the number of users (i.e., devices) involved in the collaboration process.
- *Decisions "on the fly"*: Today, most of people have a smartphone. The use of these devices by the experts and patients allows that someone with one of these devices can interact with other users and make decisions at any time and anywhere.
- *Flexibility*: Most of the cloud solutions are based on the service concept. The use of Service Oriented Architectures (SOA) enables the possibility to add new services (functionalities) easily.
- *Centralized information*: The centralized information enables that different users can access to the same information from different devices. That means, for example, that an expert can decide to take decisions through different devices.

The cloud architecture (based on a Service Oriented Architecture) proposed (Figure 4), provides a set of services about different concerns to ensure to address the features presented in the collaborative scenario. The API of the architecture should be different in each system, since usually, the solutions are customized to solve a specific problem, but the coordination, which provides the interaction and collaboration, should be common for all systems supported by the solution presented in this paper.

The set of services that manages notifications supports the interchange of push notifications between users. These services internally interact with the coordinator (i.e., a specific software component), which manages the events before sending each notification, in order to coordinate the events/notifications between the different types of users. Internally, the coordinator manages the different types of notifications and the state.

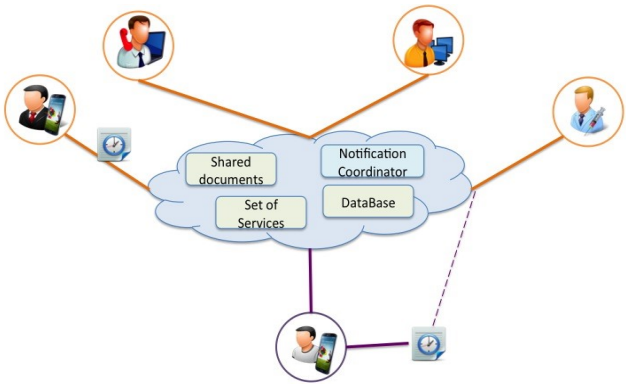

Fig. 4. Sharing informating and services through the cloud

Furthermore, it is possible that the solution proposed to support collaborative scenarios has unforeseen requirements. In this case, the architecture is based on SOA principles, so that new services could be added and new interfaces (API) could be designed in order to support different situations such as:

- *Different types of experts*: Some solutions will be designed to be used by a coach and a nutritionist, while others could be designed to be used by a wide range of experts from different domains. The architecture should support these different types of experts.
- *Different types of end-users*: Although the collaborative scenario proposes one type of end-user, it is possible that some solution proposes different types of end-users according to some characteristics or according to the activity performed (e.g., athletes, patients, etc.).
- *Custom notifications*: The collaborative scenario (Figure 1) shows the basic notifications to ensure the interaction/communication between the different types of users of the systems, but new type of notifications may be needed to cover some situations or scenarios.

## IV. APPLICATION EXAMPLE

In order to validate the proposal presented in this paper, this section presents a wellness application named CloudFit, which is focused on monitoring the health condition and wellness of athletes in Spain. Although there is already a functional prototype of this application, it is still being developed in collaboration with experts in Sport Science. The system is based on cloud technology, wearable devices (e.g., heart rate sensors, accelerometers, etc.) and smartphones; and it is made up of a Web platform and a smartphone application.

Provided that CloudFit has been designed to support athletes, it involves various kinds of users; for instance, experts from several domains (e.g., coaches, nutritionists, physicians, etc.) and the athletes (according to the collaboration scenarios presented in Figure 1). The supervisors (experts) use the Web platform to design the trainings, diets and other kinds of tasks for each athlete. This latter performs these tasks and reports the results using the smartphone application (Figure 5). The information is uploaded to the cloud to make it accessible to the experts.

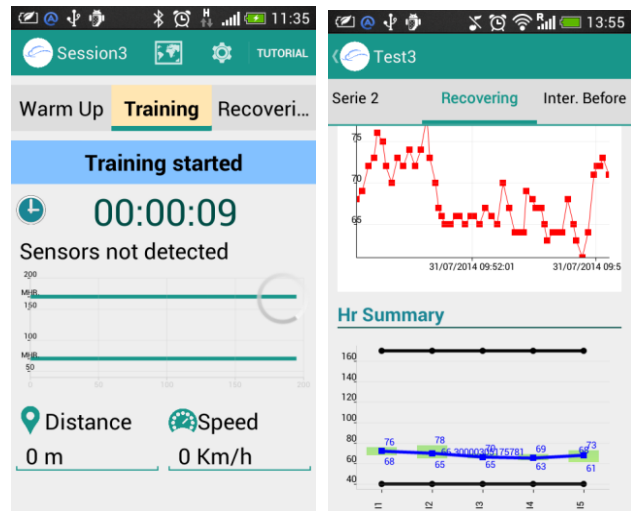

Fig. 5. User interface of the monitoring system

The collaborative work model and the architecture presented in this paper support CloudFit. The end-users receive and send notifications from/to the experts/supervisors (Figure 6) through the cloud.

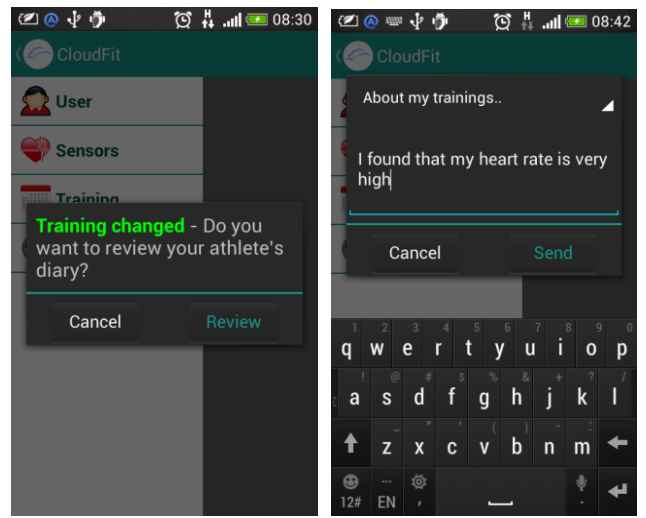

Fig. 6. User interface of CloudFit for exchanging notifications

## V. CONCLUSIONS AND FUTURE WORK

Nowadays, wellbeing is an importance concern at all levels in the society. The emergence of mobile technologies, such as smartphones, and computer paradigms, such as cloud computing, has brought about the appearance of many hand-held activity monitoring systems and applications for this domain. Most of these solutions are used by people as a tracker or diary (to know about his/her evolution) and are used by the users on their own, with no supervision of experts.

Although these solutions could be useful for many people, they are not suitable for people with special healthcare needs or patients suffering from chronic diseases. In those cases, the supervision of an expert to monitor the evolution of the patients and their goals, in terms of healthcare condition, is mandatory.

Because of this, novel sophisticated systems have appeared, supporting the activities of the supervisors and the patients.

In some contexts, the supervision by a specific expert in a particular domain is not enough to ensure the well being of the patient, since many aspects related to different healthcare domains could be involved in his/her treatment. Furthermore, the collaboration between different kinds of experts to ensure the proper evolution of the patient is a crucial concern.

In this paper a solution that intends to address the stated challenges is presented. The solution is based on a collaborative work model that supports the proper collaboration in a wellness scenario, and system architecture to support this model. This model is currently applied in a wellness platform, called CloudFit, which is currently being developed in collaboration with Sport Science practitioners and researchers for monitoring athletes in Spain.

We are currently working in the improvement of the collaborative work model, to ensure that it covers a broad spectrum of requirements in healthcare domain, as well as in the development of the CloudFit platform. We also plan to validate the proposal from the feedback obtained of Sport Science researchers through different satisfaction and quality of experience interviews.

ACKNOWLEDGMENTS

This work was partially supported by Fondecyt (Chile), Grant Nº 1120207. The authors would also like to acknowledge the time and input of Sport Science researchers and collaborators from the Instituto Mixto Universitario de Deporte y Salud (IMUDS) at the University of Granada.